\documentclass[aps,prl,superscriptaddress,showpacs,floatfix,twocolumn]{revtex4}

\usepackage{amsmath}

  \usepackage[dvips]{graphicx}

\usepackage[colorlinks=true,linkcolor=blue,plainpages=false]{hyperref}
\def\pT{\mbox{$p_T$ }}
\def\v2{\mbox{$v_2$}}
\def\pTRef{\mbox{$p_{T,{ref}}$}}

\def\sqrtsNN{\mbox{$\sqrt{s_{NN}}$}}

\newcommand{\mean}[1]{\left\langle #1 \right\rangle} 
 
\newcommand{\cumul}[1]{\left\langle\!\!\left\langle #1 
   \right\rangle\!\!\right\rangle}

\begin{document}

\newcommand{\abilene}{Abilene Christian University, Abilene, TX 79699, USA}
\newcommand{\acadsin}{Institute of Physics, Academia Sinica, Taipei 11529, Taiwan}
\newcommand{\banaras}{Department of Physics, Banaras Hindu University, Varanasi 221005, India}
\newcommand{\barc}{Bhabha Atomic Research Centre, Bombay 400 085, India}
\newcommand{\bnl}{Brookhaven National Laboratory, Upton, NY 11973-5000, USA}
\newcommand{\caucr}{University of California - Riverside, Riverside, CA 92521, USA}
\newcommand{\ciae}{China Institute of Atomic Energy (CIAE), Beijing, People's Republic of China}
\newcommand{\cns}{Center for Nuclear Study, Graduate School of Science, University of Tokyo, 7-3-1 Hongo, Bunkyo, Tokyo 113-0033, Japan}
\newcommand{\colorado}{University of Colorado, Boulder, CO 80309}
\newcommand{\columbia}{Columbia University, New York, NY 10027 and Nevis Laboratories, Irvington, NY 10533, USA}
\newcommand{\dapnia}{Dapnia, CEA Saclay, F-91191, Gif-sur-Yvette, France}
\newcommand{\debrecen}{Debrecen University, H-4010 Debrecen, Egyetem t{\'e}r 1, Hungary}
\newcommand{\elte}{ELTE, E{\"o}tv{\"o}s Lor{\'a}nd University, H - 1117 Budapest, P{\'a}zm{\'a}ny P. s. 1/A, Hungary}
\newcommand{\fsu}{Florida State University, Tallahassee, FL 32306, USA}
\newcommand{\gsu}{Georgia State University, Atlanta, GA 30303, USA}
\newcommand{\hiroshima}{Hiroshima University, Kagamiyama, Higashi-Hiroshima 739-8526, Japan}
\newcommand{\ihepprot}{Institute for High Energy Physics (IHEP), Protvino, Russia}
\newcommand{\illuiuc}{University of Illinois at Urbana-Champaign, Urbana, IL 61801}
\newcommand{\isu}{Iowa State University, Ames, IA 50011, USA}
\newcommand{\jinrdubna}{Joint Institute for Nuclear Research, 141980 Dubna, Moscow Region, Russia}
\newcommand{\kaeri}{KAERI, Cyclotron Application Laboratory, Seoul, South Korea}
\newcommand{\kangnung}{Kangnung National University, Kangnung 210-702, South Korea}
\newcommand{\kek}{KEK, High Energy Accelerator Research Organization, Tsukuba-shi, Ibaraki-ken 305-0801, Japan}
\newcommand{\kfki}{KFKI Research Institute for Particle and Nuclear Physics (RMKI), H-1525 Budapest 114, POBox 49, Hungary}
\newcommand{\korea}{Korea University, Seoul, 136-701, Korea}
\newcommand{\kurchatov}{Russian Research Center ``Kurchatov Institute", Moscow, Russia}
\newcommand{\kyoto}{Kyoto University, Kyoto 606, Japan}
\newcommand{\labllr}{Laboratoire Leprince-Ringuet, Ecole Polytechnique, CNRS-IN2P3, Route de Saclay, F-91128, Palaiseau, France}
\newcommand{\lawllnl}{Lawrence Livermore National Laboratory, Livermore, CA 94550, USA}
\newcommand{\losalamos}{Los Alamos National Laboratory, Los Alamos, NM 87545, USA}
\newcommand{\lpc}{LPC, Universit{\'e} Blaise Pascal, CNRS-IN2P3, Clermont-Fd, 63177 Aubiere Cedex, France}
\newcommand{\lund}{Department of Physics, Lund University, Box 118, SE-221 00 Lund, Sweden}
\newcommand{\muenster}{Institut f\"ur Kernphysik, University of Muenster, D-48149 Muenster, Germany}
\newcommand{\myongji}{Myongji University, Yongin, Kyonggido 449-728, Korea}
\newcommand{\nagasaki}{Nagasaki Institute of Applied Science, Nagasaki-shi, Nagasaki 851-0193, Japan}
\newcommand{\newmex}{University of New Mexico, Albuquerque, NM 87131, USA }
\newcommand{\nmsu}{New Mexico State University, Las Cruces, NM 88003, USA}
\newcommand{\ornl}{Oak Ridge National Laboratory, Oak Ridge, TN 37831, USA}
\newcommand{\orsay}{IPN-Orsay, Universite Paris Sud, CNRS-IN2P3, BP1, F-91406, Orsay, France}
\newcommand{\peking}{Peking University, Beijing, People's Republic of China}
\newcommand{\pnpi}{PNPI, Petersburg Nuclear Physics Institute, Gatchina, Russia}
\newcommand{\riken}{RIKEN (The Institute of Physical and Chemical Research), Wako, Saitama 351-0198, JAPAN}
\newcommand{\rikjrbrc}{RIKEN BNL Research Center, Brookhaven National Laboratory, Upton, NY 11973-5000, USA}
\newcommand{\rikkyo}{Physics Department, Rikkyo University, 3-34-1 Nishi-Ikebukuro, Toshima, Tokyo 171-8501, Japan}
\newcommand{\saispbstu}{St. Petersburg State Technical University, St. Petersburg, Russia}
\newcommand{\saopaulo}{Universidade de S{\~a}o Paulo, Instituto de F\'{\i}sica, Caixa Postal 66318, S{\~a}o Paulo CEP05315-970, Brazil}
\newcommand{\seoulnat}{System Electronics Laboratory, Seoul National University, Seoul, South Korea}
\newcommand{\stonybrkc}{Chemistry Department, Stony Brook University, Stony Brook, SUNY, NY 11794-3400, USA}
\newcommand{\stonycrkp}{Department of Physics and Astronomy, Stony Brook University, SUNY, Stony Brook, NY 11794, USA}
\newcommand{\subatech}{SUBATECH (Ecole des Mines de Nantes, CNRS-IN2P3, Universit{\'e} de Nantes) BP 20722 - 44307, Nantes, France}
\newcommand{\tenn}{University of Tennessee, Knoxville, TN 37996, USA}
\newcommand{\titech}{Department of Physics, Tokyo Institute of Technology, Tokyo, 152-8551, Japan}
\newcommand{\tsukuba}{Institute of Physics, University of Tsukuba, Tsukuba, Ibaraki 305, Japan}
\newcommand{\vandy}{Vanderbilt University, Nashville, TN 37235, USA}
\newcommand{\waseda}{Waseda University, Advanced Research Institute for Science and Engineering, 17 Kikui-cho, Shinjuku-ku, Tokyo 162-0044, Japan}
\newcommand{\weizmann}{Weizmann Institute, Rehovot 76100, Israel}
\newcommand{\yonsei}{Yonsei University, IPAP, Seoul 120-749, Korea}
\affiliation{\abilene}
\affiliation{\acadsin}
\affiliation{\banaras}
\affiliation{\barc}
\affiliation{\bnl}
\affiliation{\caucr}
\affiliation{\ciae}
\affiliation{\cns}
\affiliation{\colorado}
\affiliation{\columbia}
\affiliation{\dapnia}
\affiliation{\debrecen}
\affiliation{\elte}
\affiliation{\fsu}
\affiliation{\gsu}
\affiliation{\hiroshima}
\affiliation{\ihepprot}
\affiliation{\illuiuc}
\affiliation{\isu}
\affiliation{\jinrdubna}
\affiliation{\kaeri}
\affiliation{\kangnung}
\affiliation{\kek}
\affiliation{\kfki}
\affiliation{\korea}
\affiliation{\kurchatov}
\affiliation{\kyoto}
\affiliation{\labllr}
\affiliation{\lawllnl}
\affiliation{\losalamos}
\affiliation{\lpc}
\affiliation{\lund}
\affiliation{\muenster}
\affiliation{\myongji}
\affiliation{\nagasaki}
\affiliation{\newmex}
\affiliation{\nmsu}
\affiliation{\ornl}
\affiliation{\orsay}
\affiliation{\peking}
\affiliation{\pnpi}
\affiliation{\riken}
\affiliation{\rikjrbrc}
\affiliation{\rikkyo}
\affiliation{\saispbstu}
\affiliation{\saopaulo}
\affiliation{\seoulnat}
\affiliation{\stonybrkc}
\affiliation{\stonycrkp}
\affiliation{\subatech}
\affiliation{\tenn}
\affiliation{\titech}
\affiliation{\tsukuba}
\affiliation{\vandy}
\affiliation{\waseda}
\affiliation{\weizmann}
\affiliation{\yonsei}
\author{S.S.~Adler}	\affiliation{\bnl}
\author{S.~Afanasiev}	\affiliation{\jinrdubna}
\author{C.~Aidala}	\affiliation{\bnl} \affiliation{\columbia}
\author{N.N.~Ajitanand}	\affiliation{\stonybrkc}
\author{Y.~Akiba}	\affiliation{\kek}  \affiliation{\riken}  \affiliation{\rikjrbrc}
\author{A.~Al-Jamel}	\affiliation{\nmsu}
\author{J.~Alexander}	\affiliation{\stonybrkc}
\author{R.~Amirikas}	\affiliation{\fsu}
\author{K.~Aoki}	\affiliation{\kyoto} \affiliation{\riken}
\author{L.~Aphecetche}	\affiliation{\subatech}
% ...
\collaboration{PHENIX Collaboration} \noaffiliation

\hyphenation{author another created financial paper re-commend-ed Post-Script}

\title{ Saturation of azimuthal anisotropy in Au~+~Au collisions 
at~$\sqrt{s_{NN}}$~=~62~-~200~GeV }

\begin{abstract}

New measurements are presented for charged hadron azimuthal correlations
at mid-rapidity in Au+Au collisions at \sqrtsNN~=~62.4 and 200~GeV. They
are compared to earlier measurements obtained at \sqrtsNN~=~130~GeV and in
Pb+Pb collisions at \sqrtsNN~=~17.2~GeV. Sizeable anisotropies are
observed with centrality and transverse momentum ($p_T$) dependence
characteristic of elliptic flow ($v_2$).  For a broad range of
centralities, the observed magnitudes and trends of the differential
anisotropy, \v2($p_T$), change very little over the collision energy range
\sqrtsNN~=~62~-~200~GeV, indicating saturation of the excitation function
for $v_2$ at these energies. Such a saturation may be indicative of the
dominance of a very soft equation of state for \sqrtsNN~$\sim$~60~-~200~GeV.

\end{abstract}
\maketitle


	Extremely high energy-density nuclear matter is 
produced in energetic Au+Au collisions at the 
Relativistic Heavy Ion Collider (RHIC)~\cite{QM:2004,Nagle:2004fe}. 
The dynamical evolution of this matter is predicted to reflect the presence and 
evolution of the Quark Gluon Plasma (QGP) -- a new 
phase of nuclear matter~\cite{Gyulassy:2004zy,Muller:2004kk,Shuryak:2004cy}. 
Azimuthal correlation measurements are important in several ways.
They serve as a ``barometric sensor" for pressure gradients developed in the 
collision and hence yield insight into crucial issues of thermalization 
and the equation of state 
(EOS)~\cite{Ollitrault:1992bk,Kolb:2001qz,Hirano:2004rs}.
They provide important constraints for the density of the medium and the 
effective energy loss of partons which traverse 
it~\cite{Gyulassy:2000gk,Molnar:2001ux}.
They can provide valuable information on the gluon saturation scale in the 
nucleus~\cite{Kovchegov:2002nf}.

	Recent measurements at RHIC (\sqrtsNN~=~130 and 200~GeV) indicate 
a mixture of (di-)jet and harmonic contributions to azimuthal 
correlations in Au+Au collisions~\cite{Ajitanand:2002qd,Chiu:2002ma,Adler:2002ct,Adler:2002tq,Adler:2004zd}. 
The asymmetric (di-)jet contributions are found to be relatively small but can be 
separated; they show an increase with \pT and indicate strong suppression 
of away-side jet yields~\cite{Adler:2002tq}. Significant modifications to the 
away-side jet topology have also been reported~\cite{Rak:2004gk}.
These observations, which are particularly
striking for very central collisions, have been interpreted as evidence for 
parton energy loss and jet quenching in the produced 
medium~\cite{Gyulassy:2004zy}. 
	The harmonic contributions show significant strength at mid-rapidity 
with characteristic dependencies on 
\pT and centrality~\cite{Ajitanand:2002qd,Adcox:2002ms,Adler:2002pu,Adler:2003kt}.
They are typically characterized by the second order Fourier coefficient, 
%
 $ v_2 = \mean{e^{i2(\phi_1-\Phi_{RP})}} $,
%
where $\phi_{1}$ represents the azimuthal emission angle of a charged 
hadron and $\phi_{RP}$ is the azimuth of the reaction 
plane. The brackets denote statistical averaging over particles and events.  
At low \pT  ($p_T \mathbin{\lower.3ex\hbox{$\buildrel<\over
{\smash{\scriptstyle\sim}\vphantom{_x}}$}} 2.0$ GeV/$c$)
the magnitude and trends of \v2 are under-predicted by hadronic cascade 
models supplemented with string 
dynamics~\cite{Bleicher:2000sx}, 
but are well reproduced by models which incorporate hydrodynamic 
flow~\cite{Shuryak:2004cy,Kolb:2001qz}.
This has been interpreted as evidence for the production of a 
thermalized state of partonic matter~\cite{Gyulassy:2004zy,Muller:2004kk,Shuryak:2004cy}. 
At higher \pT~the predictions of quark coalescence~\cite{Fries:2003kq}
are consistent with the data~\cite{Adler:2003kt,Adams:2003am}, and quantitative 
agreement has been achieved with transport model calculations which incorporate 
large opacities~\cite{Molnar:2001ux}.

	At Super Proton Synchrotron (SPS) energies (\sqrtsNN~$\sim 17$ GeV)
azimuthal correlation measurements also indicate a mixture of 
(di-)jet and harmonic contributions~\cite{Agakichiev:2003gg,Alt:2003ab}. However, 
the observed anisotropy of the harmonic contribution is approximately 50\% of the value
observed at full RHIC energy (\sqrtsNN~=~200~GeV). Therefore, an important outstanding 
issue is the detailed behavior of \v2 over the range which spans 
SPS~-~RHIC energies. In recent work, the PHOBOS collaboration has investigated the 
patterns for $p_T$-integrated \v2 over a broad range of pseudorapidities~\cite{phobos:2004zg}. 
Here, we present more revealing differential measurements for Au~+~Au collisions 
at \sqrtsNN~=~62.4~-~200 GeV and the first excitation function for differential \v2 
which spans beam energies from the Alternating Gradient Synchrotron (AGS) to 
RHIC (\sqrtsNN~$\sim$~3~-~200 GeV).

 	The colliding Au beams (\sqrtsNN~=~62.4, 130, and 200~GeV) 
used in the measurements presented here have been provided by RHIC in 
three separate experimental running periods (in 2000-2004). 
Charged tracks were detected in the east and west 
central arms of PHENIX~\cite{Adcox:2003zm}, 
each of which subtend 90$^\circ$ 
in azimuth $\phi$, and $\pm 0.35$ units of pseudo-rapidity $\eta$. 
Track reconstruction was accomplished at each collision energy via pattern recognition using 
a drift chamber (DC) followed by two layers of multi-wire proportional
chambers with pad readout (PC1, PC3)~\cite{Adcox:2003zm}. A combinatorial 
Hough transform in the track bend plane was used for pattern recognition 
in the DC~\cite{Mitchell:2002wu}. 
For each analysis, the collision vertex $z$ along the beam direction was 
constrained to be within $|z| <$~30~cm.  A confirmation hit within a $2\sigma$ 
matching window was required in PC3, located at a radius 
of 5 m, to eliminate most albedo, conversions, and decays. 
Particle momenta were measured with resolutions $\delta p/p = 0.7\% \oplus 0.91\%\,p$, 
$\delta p/p = 0.6\% \oplus 3.6\%\,p$, and $\delta p/p = 0.7\% \oplus 1.0\%\,p$~(GeV/$c$) 
at \sqrtsNN~=~62.4, 130, and 200~GeV respectively, good enough to have very little 
influence, if any, on the results presented here. 

	Event centralities were obtained at \sqrtsNN~=~62.4 GeV via a series of cuts on the 
analog response of the PHENIX beam counters (BBC). For \sqrtsNN~=~130 and 200 GeV, cuts 
in the space of BBC versus ZDC analog response were employed; 
they reflect percentile cuts on the total 
interaction cross section at each beam energy \cite{Adcox:2003nr}. 
Estimates for the number of participant nucleons N$_{part}$, were 
also made for each of these cuts following the Glauber-based 
model detailed in Ref.~\cite{Adcox:2003nr}. Systematic uncertainties associated with 
these determinations are estimated to be less than $\sim 10$\% for 
central and mid-central collisions

The differential \v2 measurements reported in this Letter have been obtained
via three separate methods of analysis. In the first, we used the reaction plane  
technique which correlates the azimuthal angles of charged tracks detected in the central arms
with the azimuth of an estimated event plane $\Phi_2$, determined via hits in the North and South
BBC's located at $\mid{\eta}\mid \sim 3 - 3.9$~\cite{Adler:2003kt}. This method 
was used for the analysis of data taken at both \sqrtsNN~=~62.4 and 200~GeV. 
Corrections~\cite{Poskanzer:1998yz,Adler:2003kt} were applied to account for possible 
azimuthal distortions in the distribution of the estimated reaction planes.
Values of \v2 were calculated via the expression
\[
\v2 = {\frac{\left\langle \cos(2(\phi -\Phi_2))\right\rangle}{\left\langle\cos(2(\Phi_2-\Phi_{RP}))\right\rangle}},
\]
where the denominator represents a resolution factor which corrects for
the difference between the estimated and the true azimuth of the 
reaction plane $\Phi_{RP}$~\cite{Poskanzer:1998yz,Adler:2003kt}. The estimated resolution of the combined reaction 
plane from both BBC's~\cite{Adler:2003kt} has an average of 0.33 (0.16) over centrality with 
a maximum of about 0.42 (0.19) for \sqrtsNN=~200~(62.4)~GeV. 
Thus, the estimated correction factor, which is the inverse of the resolution 
for the combined reaction plane, ranges from 2.4 (5.4) to 5.0 (13).

	In the second method, a cumulant analysis
was performed on data collected at \sqrtsNN~=~200 and 62.4~GeV to obtain 
the anisotropy directly~\cite{Borghini:2001vi}
\begin{equation}
\mean{e^{2i(\phi_1-\phi_2)}}=
\mean{e^{2i\phi_1}}\mean{e^{-i2\phi_2}} +
\cumul{e^{2i(\phi_1-\phi_2)}}. 
\label{eq2}
\end{equation}
Here, the double brackets denote an average over pairs of particles emitted in 
an event followed by further averaging over events. 
For a detector having full azimuthal acceptance, the averages 
$\mean{e^{2i\phi_1}}$ and $\mean{e^{-2i\phi_2}}$ vanish due to  
symmetry considerations, to give the second order cumulant estimate
$\v2\{2\}$~\cite{Borghini:2001vi} of \v2 
\begin{equation}
\cumul{e^{2i(\phi_1-\phi_2)}} = \v2\{2\}^2.
\label{2nd_cum}
\end{equation}
Since PHENIX does not have full azimuthal acceptance, 
$\mean{e^{2i\phi_1}}$ and $\mean{e^{-2i\phi_2}}$ do not vanish and this 
leads to an initial underestimate of the extracted anisotropy. 
To correct for this underestimate, separate correction factors ($\sim 30\%$) 
were evaluated and applied for each centrality and \pT cut, at each collision energy, following the 
procedures detailed in Ref.~\cite{Borghini:2001vi}. 

	In the third method, we extracted the anisotropy at \sqrtsNN~=~62.4, 130 and 200~GeV via 
assorted two-particle correlation functions~\cite{Adler:2003kt,Ajitanand:2002qd}:
\begin{equation}
C(\Delta\phi) = \frac{N_{cor}(\Delta\phi)}{N_{mix}(\Delta\phi)}. 
\label{ratio}
\end{equation}
Here, $N_{cor}(\Delta\phi)$ is the observed $\Delta\phi$ distribution
for charged particle pairs selected from the same event,  
and $N_{mix}(\Delta\phi)$ is the $\Delta\phi$ distribution for particle 
pairs selected from mixed events. Mixed events were obtained by randomly 
selecting each member of a particle pair from different events  
with the same multiplicity and vertex cuts.

	To extract the anisotropy of these correlations, two 
correlation functions were generated for each \pT and centrality 
selection~\cite{Adler:2003kt,Ajitanand:2002qd}. For the first, charged hadron 
pairs were formed by selecting both particles 
from a reference range \pTRef, which excluded the \pT range of interest 
(i.e. a reference correlation). For the second, assorted hadron 
pairs were formed by selecting one member from the \pT-range of interest and 
the other from \pTRef. The elliptic flow \v2, was obtained via the ratio
$A_{2,a}$/$\surd{A_{2,{ref}}} = \v2$ where $A_{2,a}$ and $A_{2,{ref}}$ are 
the anisotropies extracted from the assorted and reference correlation 
functions (respectively) with the fit function: 
\begin{equation}
C(\Delta \phi)= a_1\left(1+ 2A_{2}cos(2\Delta \phi) + \lambda e^{(-0.5(\Delta \phi/\sigma)^2)} \right)
\label{FitFunction}
\end{equation}
where the Gaussian and harmonic terms are used to characterize the 
asymmetry (at small $\Delta \phi$) and the anisotropy of the correlation function respectively~\cite{Ajitanand:2002qd,Adler:2002ct,Adler:2002tq}. 

%
\begin{figure}[tbh]
\includegraphics[width=1.0\linewidth]{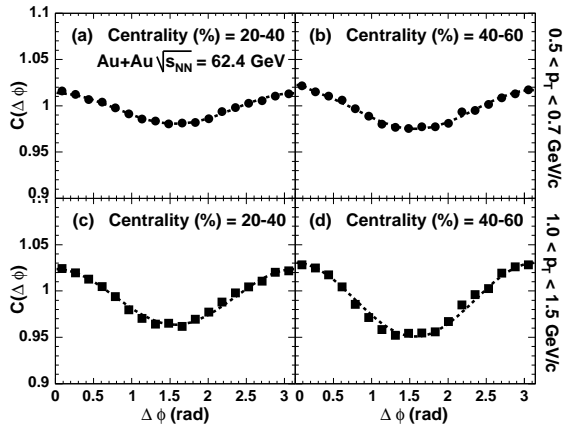}
\caption{ \label{fig1}
	Assorted-\pT correlation functions ($0.65 < p_{T,ref} < 2.5$~GeV/c) for charged 
hadrons of $0.5<p_T<0.7$ GeV/$c$ (top panels) and $1.0<\pT<1.5$ (bottom panels) obtained in 
Au+Au collisions at \sqrtsNN~=~62.4 GeV. The left and right panels show correlation 
functions for centrality cuts of 20-40\% and 40-60\% respectively. 
The lines represent fits to the correlation functions (see text). 
}
\end{figure}

Figures \ref{fig1}a - \ref{fig1}d show representative $\Delta\phi$ correlation
functions obtained for charged hadrons detected in the PHENIX central 
arms ($-0.35 < \eta < 0.35$) at \sqrtsNN~=~62.4~GeV. 
Correlation functions for mid-central
events (centrality = 20~-~40\%)  are shown for hadrons
with $0.5 < p_T <0.7$ GeV/$c$ and $1.0 < p_T <1.5$ GeV/$c$ in
Figs.~\ref{fig1}a and c respectively.  The same $p_T$ cuts have been
made for the correlation functions shown in Figs.~\ref{fig1}b and d but
for more peripheral collisions (centrality = 40~-~60\%). For both sets 
of correlation functions $0.65 < p_{T,ref} < 2.5$~GeV/c. 
Figs.~\ref{fig1}a - \ref{fig1}d show a clear anisotropic pattern with relatively
small asymmetries ($0^{\rm o}/180^{\rm o}$ ratios). 
Such asymmetries have been attributed to small jet contributions to the 
correlation functions~\cite{Ajitanand:2002qd,Adler:2002tq}, 
and are expected to decrease with decreasing \sqrtsNN. 
The curves in Fig.~\ref{fig1} indicate a fit to the correlation 
function with Eq.~\ref{FitFunction}; they show 
an increase of the anisotropy with increasing impact 
parameter and $p_T$. These trends are similar to those of prior AGS, SPS and RHIC 
measurements~\cite{Chung:2001qr,Agakichiev:2003gg,Alt:2003ab,Adcox:2002ms} and are 
consistent with the expected patterns 
for in-plane elliptic flow~\cite{Shuryak:2004cy,Kolb:2001qz}. 
%

%
\begin{figure}[bth]
\includegraphics[width=1.0\linewidth]{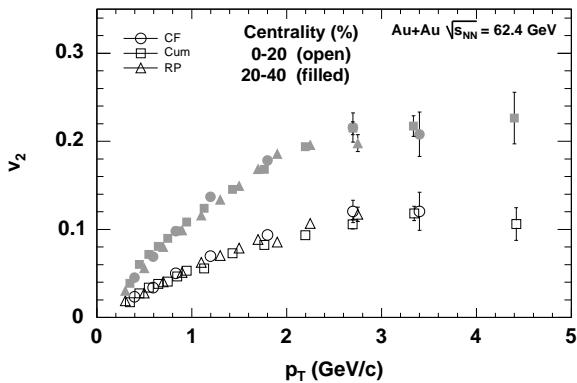}
\caption{ \label{fig2}
Differential anisotropy $v_2(p_T)$ for charged hadrons in Au+Au  
collisions at $\sqrt{s_{NN}}$ = 62.4 GeV with centrality cuts of 0-20\%  
(open symbols) and 20-40\% (filled symbols), obtained via the methods of  
correlation functions (CF), cumulants (Cum) and reaction plane (RP).
}
\end{figure}

Figure~\ref{fig2} compares the differential anisotropy \v2($p_T$), 
obtained at \sqrtsNN~=~62.4 GeV for all three methods of extraction. The 
error bars shown indicate statistical errors. Systematic errors are estimated 
to be $\sim$~10\%, 5\%, and 5\% for extractions via the reaction plane, cumulant 
and correlation function methods of analysis respectively. The results, which are 
shown for two separate centrality cuts (0~-~20\% and 20~-~40\%) in each case, 
indicate an initial increase of \v2 with  
\pT followed by the previously observed plateau for  
\pT $\mathbin{\lower.3ex\hbox{$\buildrel>\over
{\smash{\scriptstyle\sim}\vphantom{_x}}$}}$ 
2.5 GeV/$c$~\cite{Adler:2002pu,Ajitanand:2002qd}. 
The close agreement of \v2($p_T$) values obtained from the cumulant and 
correlation function methods of analysis, serve to confirm the reliability of these 
methods of extraction. On the other hand, the agreement between results from these 
latter methods and that obtained from the reaction plane method is quite striking, 
given the large rapidity gap ($\sim 3$ units) between the particles used for 
reaction plane determination and the mid-rapidity particles
correlated with this plane. It is expected that the latter correlations are less 
influenced by non-flow contributions especially for $p_T < 2.0$~GeV/$c$.
Consequently, we attribute this agreement to the absence of  
strong non-flow contributions to the hadron correlations 
(for $\pT < 2.0$~GeV/$c$) at mid-rapidity. A similarly good agreement between 
the different methods of analysis was obtained for 
all centralities presented in this work. 


%
\begin{figure}[tbh]
\includegraphics[width=1.0\linewidth]{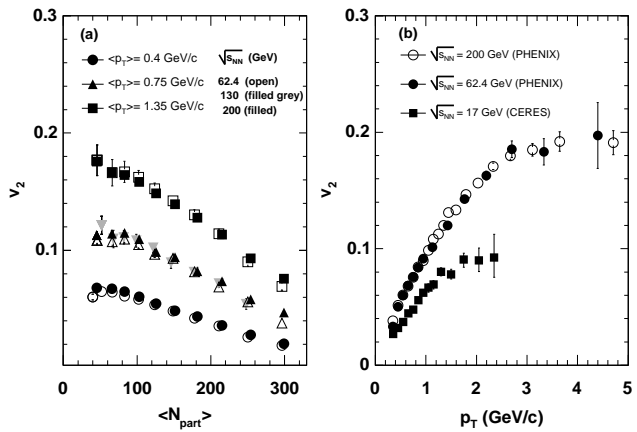}
\vspace*{-0.02in}
\caption{ \label{fig3} 
	Differential anisotropy \v2(N$_{part}$) (left) 
and \v2($p_T$) (right) for several energies as indicated. \v2($p_T$) is 
shown for the centrality selection 13~-~26\%. The CERES data are 
taken from Ref.~\cite{Agakichiev:2003gg}. 
}
\end{figure}

Figures~\ref{fig3}a and~\ref{fig3}b compare the centrality and $p_T$ dependence  
(respectively) of the anisotropy obtained at several collision energies. 
The circles, stars and squares in \ref{fig3}a show $v_2$(N$_{part}$) 
for $\langle p_T \rangle$ selections of 0.4, 0.75 and 1.35 GeV/$c$ obtained 
via the cumulant and correlation function methods of analysis. The same results
obtained via the reaction plane method are consistent with prior results\cite{Adler:2003kt}. 
The open and filled symbols show measurements performed 
at \sqrtsNN~=~62.4 and 130~(200) GeV as indicated; they show rather 
striking agreement between the magnitudes of the \v2 values obtained at 
all three collision energies. Further evidence that this agreement persists 
down to \sqrtsNN~=~62.4 GeV is given in Fig.~\ref{fig3}b. 
Here, the open and filled circles compare the differential 
anisotropy $v_2(p_T)$, obtained at \sqrtsNN~=~62.4 and 200~GeV 
for the 13-26\% most central collisions. The comparison indicates 
little change in \v2 as the collision energy is raised 
from \sqrtsNN~=~62.4 to 200 GeV. This contrasts with the much lower $v_2$ values 
measured in Pb+Pb collisions (filled squares) by the CERES collaboration 
at $\sqrt{s_{NN}}$ = 17.2 GeV, for the same centrality 
cut (13~-~26\%)~\cite{Agakichiev:2003gg}.

%
\begin{figure}[tbh]
\includegraphics[width=1.0\linewidth]{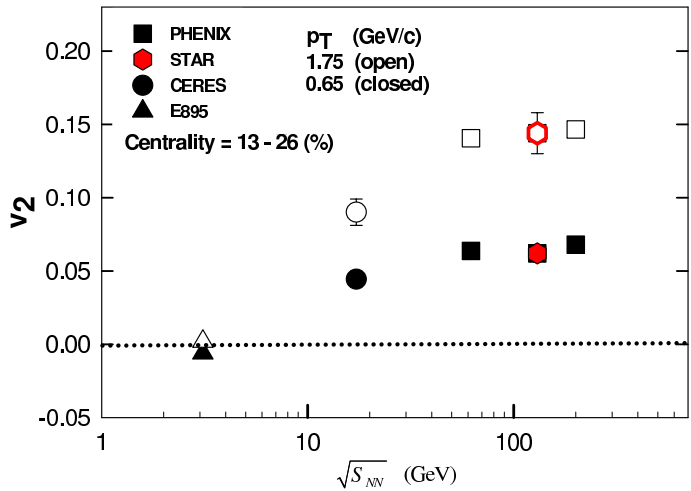}
\caption{ \label{fig4} 
	Differential \v2~vs.~\sqrtsNN~for charged hadrons in nucleus-nucleus collisions. 
Results are shown for the centrality cut of 13~-~26\% and \pT~selections 
of 1.75 GeV/$c$ (open symbols) and 0.65 GeV/$c$ (closed symbols). The STAR,
CERES and E895 data are taken from Refs.~\cite{Adler:2002pu}, \cite{Agakichiev:2003gg}
and~\cite{Chung:2001qr,Danielewicz:2002pu,Pinkenburg:1999ya} respectively.
}
\end{figure}

The \sqrtsNN~dependence of \v2 for charged hadrons produced in Au+Au 
collisions is summarized  in Fig.~\ref{fig4} for two separate \pT selections 
(0.65 and 1.75 GeV/$c$) and centrality = 13-26\%. These data are 
taken from the current measurements and earlier measurements at 
the SPS~\cite{Agakichiev:2003gg} and the 
AGS~\cite{Chung:2001qr,Danielewicz:2002pu,Pinkenburg:1999ya}. 
The AGS measurements~\cite{Chung:2001qr,Danielewicz:2002pu,Pinkenburg:1999ya} are for protons.
The STAR results were obtained for a slightly different centrality selection  
(10-30\%)~\cite{Adler:2002pu} having essentially the same mean centrality.
For both \pT~cuts, the magnitude of \v2 shows a significant increase with 
collision energy ($\sim$~50\% increase from SPS to RHIC) up to the energy 
\sqrtsNN~=~62.4~GeV. Thereafter, it appears to saturate for larger 
beam energies.

	To summarize, we have measured differential azimuthal 
anisotropies for charged hadrons in Au~+~Au collisions spanning the energy 
range \sqrtsNN~=~62.4~-~200 GeV. Detailed comparisons of these differential 
measurements indicate no significant collision energy dependence of the anisotropy 
over this range. By contrast, comparisons to differential measurements obtained at AGS 
and SPS energies indicate that \v2 increases with collision energy up 
to \sqrtsNN~=~62.4~GeV. Given the fact that the energy density is estimated to 
increase by approximately 30\% over the range \sqrtsNN~=~62.4~-~200 GeV,  
this apparent saturation of \v2 above \sqrtsNN~=~62.4 GeV may be 
indicative of the role of a rather soft equation of state. Such a softening could 
result from the production of a mixed phase~\cite{Danielewicz:2002pu} for the 
range \sqrtsNN~=~62.4~-~200 GeV. 


We thank the staff of the Collider-Accelerator and Physics
Departments at BNL for their vital contributions.  
We acknowledge support from the 
Department of Energy and NSF (U.S.A.), 
MEXT and JSPS (Japan), 
CNPq and FAPESP (Brazil), 
NSFC (China), 
IN2P3/CNRS, and CEA (France), 
BMBF, DAAD, and AvH (Germany), 
OTKA (Hungary), 
DAE and DST (India), 
ISF (Israel), 
KRF, CHEP, and KOSEF (Korea), 
RMIST, RAS, and RMAE (Russia), 
VR and KAW (Sweden), 
U.S. CRDF for the FSU, 
US-Hungarian NSF-OTKA-MTA, 
and US-Israel BSF.


\bibliographystyle{apsrev}
\bibliography{ppg047x0} 

\end{document}